# Wireless-Delimited Secure Zones with Encrypted Attribute-Based Broadcast for Safe Firearms

Marcos Portnoi    Chien-Chung Shen
Department of Computer and Information Sciences
University of Delaware
Newark, DE, U.S.A.
{mportnoi, cshen}@udel.edu

*Abstract*—This work presents an application of the highly expressive Attribute-Based Encryption to implement wireless-delimited Secure Zones for firearms. Within these zones, radio-transmitted local policies based on attributes of the consumer and the firearm are received by embedded hardware in the firearms, which then advises the consumer about safe operations. The Secure Zones utilize Attribute-Based Encryption to encode the policies and consumer or user attributes, and providing privacy and security through it cryptography. We describe a holistic approach to evolving the firearm to a cyber-physical system to aid in augmenting safety. We introduce a conceptual model for a firearm equipped with sensors and a context-aware software agent. Based on the information from the sensors, the agent can access the context and inform the consumer of potential unsafe operations. To support Secure Zones and the cyber-physical firearm model, we propose a Key Infrastructure Scheme for key generation, distribution, and management, and a Context-Aware Software Agent Framework for Firearms.

I. INTRODUCTION

Advances of digital technologies, as they are incorporated into devices for people's usage, often result in enhancements including ease of use, safety, precision, optimizations in resource consumption and costs. In particular, these technological advances spawned a class of systems named Cyber-Physical Systems, in which computational elements integrate with physical objects, capable of evolving a purely mechanical device into a cyber-aided version.

The trend has touched several areas, producing machines heavily dependent on cyber systems; onto these machines we rely, and to which we happily entrust our lives. Commercial aircraft possess a multitude of onboard systems to aid operation, from fly-by-wire components, to autopilots that follow navigation maps and can automate landing procedures. Military aircraft, such as the B-2 Stealth Bomber, have exceptionally complex avionics, without which the aircraft would be unstable to fly [1]. Automobiles usually integrate anti-lock brakes, safety airbags, and may be available with traction control and collision avoidance systems. During 1991 and 1993 seasons, the Formula One (F1) team Williams presented a car equipped with active suspension. The suspension, together with telemetry, was capable of detecting terrain configuration and self-adjusting for optimal aerodynamics and equilibrium of the car. The system gave rise to a car so vastly superior to its counterparts and so fast that further F1 regulations banned similar technological driver-aids [2]. Google is forwarding research in self-driven cars and expects them to be available in less than a decade [3].

Furthermore, patients surrender their health to pacemakers and electronic monitors. There are robotic instruments capable of performing surgery [4] (although the delicate and expensive nature of its market has recently given raise to scrutiny [5]), and ongoing research on human exoskeletons [6]. More and more, we integrate machinery with cyber-capabilities into our daily activities and experiment enhanced safety, comfort, and productivity.

One device has persisted largely untouched by cyber technology since its inception possibly in the 12$^{th}$ century: the firearm. In light of recent tragic mass shootings, communities again claim for employing technology to boost firearm safety [7], [8]. In this paper, we propose evolving this mechanical device into a cyber-physical system for greater consumer safety. In our scheme, we also combine the broadcast nature of wireless communication and the expressiveness of a one-to-many cryptosystem, Attribute-Based Encryption (ABE), to establish wireless-delimited zones within which, according to choice, a firearm can alert the consumer, or user, of unsafe operation, depending on attributes programmed in the gun's electronics. Furthermore, we discuss incorporating a range of common sensors to the firearms and enable context-aware decision capabilities; by analyzing the several inputs from the embedded sensors and evaluating context, a firearm could alert the consumer of unsafe operations by utilizing haptic and visual feedbacks.

Understandably, adding electronics to a firearm can be interpreted as adding layers of failure to a reasonably reliable piece of machinery [9]. However, as we articulated earlier, we already depend, for our safety, on engines with high degree of integration of electronics, such as aircraft and automobiles: these engines became safer and more reliable with the aid of technology. It is reasonable to expect that technology can help enhance a firearm safe operation.

In the next section, we survey related work in gun safety. In Section III, we present our view of the firearm as a cyber-physical system featuring sensors and a context-aware software agent, and in Section IV, the supporting Context-Aware Agent

Framework. We describe the fundamentals of Attribute-Based Encryption in Section V. Section VI depicts our Secure Zones and its operation. The proposed Key Infrastructure Scheme for generation, distribution and management of encryption keys is discussed in VII. We conclude the paper with future work in Section VIII.

In this document, we use the definitions "consumer" and "user" interchangeably.

## II. RELATED WORK

Integrating cyber technology in weapons, either for the purpose of increased safety or increased lethal capacity, has been proposed for many years. In particular, personal weapons which would integrate cyber systems have been known as "smart guns" [9].

Solutions that involve detecting an authorized consumer by means of a wearable radio transmitter include iGUN [10], in which the consumer wears a special ring that likely possesses an RFID transmitter. When the ring is in close proximity with the trigger, an actuator unlocks the weapon. A manual safety lock is still available, adding a second layer of security. TriggerSmart [11] employs a similar mechanism, however without the second, manual safety lock, as inferred from the available prototypes displayed in videos at the time of this writing. TriggerSmart proponents also mention "safe zones" within which guns would be remotely disabled. The guns would contain hardware based on an RF transceiver and would employ DASH7 wireless sensor protocol [12].

Armatix [13] utilizes an RFID wristwatch that, in addition, allows entering a PIN to unlock the firearm in its radio range and may use specific times for gun deactivation. A Target Response System would also deactivate the weapon if it were not on target. No details are furnished, however, on how this system operates and which sensors it involves.

The New Jersey Institute of Technology presented a prototype of firearm in which an array of transducers (more specifically, pressure sensors) distributed over the handle detects the allegedly unique way a consumer grips a firearm [14], [15]. Using this biometrics data, the weapon may recognize authorized users and activate the locking mechanism appropriately.

In our holistic approach, we augment these views by building a context-aware system in the firearm, assessing the environment from an array of sensors and constructing safety decisions. Furthermore, our model may offer different operation choices, from an off setting, to an advisory mode, to a fully locking mode, such that our solutions can be validated in short term.

## III. THE FIREARM AS CYBER-PHYSICAL SYSTEM

With current technology, a firearm could comprise a set of sensors enabling it to gather a wide range of information from the environment. A software agent inside the firearm can integrate the contextual information to form an evaluation of the immediate situation; the firearm becomes context-aware. Upon this evaluation, the firearm can advise the consumer of potential unsafe operations through visual or haptic feedback, or actuate a locking mechanism.

However, before firearms can lock themselves, society and policy/law must agree on this path. Moreover, to guarantee that guns without self-enforcement do not have an unfair advantage upon those guns that are able to self-lock, both these types of guns (ideally) cannot coexist. To overcome this difficulty in the short term, our firearm model has three modes of operation (selectable by software or by a PIN panel): *Off*, in which all cyber-systems are deactivated, and the firearm functions as a pure mechanical device; *Advisory Mode*, in which the Secure Zone and context agent act as an advisory system for safe operations, indicating unsafe situations by means of visual and haptic feedback; and *Full Lock*, in which, in addition to feedback, the actuator can operate the locking mechanism to prevent the gun from firing. Fig. 1 portrays this cyber-physical, context-aware firearm concept. This conceptual model presents the sensors, feedback outputs and their corresponding intents as listed in Table I.

TABLE I: SENSORS AND FEEDBACK IN CONCEPTUAL FIREARM MODEL.

| Sensor | Operation |
| --- | --- |
| Microphone | to detect environment audio and assess potential anomalous situations, such as abundance of noise, screams (panic?), or shouted commands. |
| Front-facing camera | to assist in detecting invalid targets (see [16] for examples of using cameras to identify gestures). |
| Rear-facing camera | can be used to identify the owner. |
| Pressure sensors in handle | used to recognize the particular grip of a consumer, or anomalous grips [15]. |
| Accelerometers | may detect excessive shaking (incompatible with proper weapon handling), among other functions. |
| Actuator-operated safety lock | if the context agent infers an anomalous behavior, or the firearm is not advised to operate according to the Secure Zone policy, an actuator locks the firearm in the Full Lock mode. |
| Haptic feedback | if the context agent infers an anomalous behavior or dangerous handling, the firearm handle can slightly vibrate or change texture, alerting the consumer of unsafe operation. |
| Context Color-coded visual feedback | in case of dangerous handling, as inferred by the context agent, a light comes on close to the user's eye of sight, alerting of unsafe operation. |
| TPD (Tamper-Proof Device) | hardware with tamper-resisting capabilities. Such device is secure, cannot be compromised and no data can be from it by adversaries [17], [18]. |
| Cyber-physical software agent | Software component of our cyber-physical firearm model, stored into the TPD. Analyzes inputs from sensors to infer context, and performs cryptography operations. The agent is build according to our Context-Aware Agent Framework, which we present in Section IV. |

Initially, we develop the necessary components for implementing the Secure Zones, described in Section VI. Those would involve radio antennas for wireless communications, the TPD with cryptography and context software agent, and feedback mechanisms.

## IV. CONTEXT-AWARE AGENT FRAMEWORK

We propose a framework for context-aware mobile applications, based on software agents (Fig. 2), applied to support the firearm cyber-physical model. These agents, implemented on firearms, act on behalf of the consumer/device. The agents interact with the environment and network through sensors, and with other agents and applicable software. From data acquired through the array of sensors, consumer input, and (possibly) other agents, the agent can build a context and analyze it using a range of artificial intelligence and statistical tools. Then, the agent can make decisions according to its goals, plans, and current state [19]. For brevity and space constraints, we do not develop a full description of this framework in this document.

## V. ATTRIBUTE-BASED ENCRYPTION

In their pivotal paper, Sahai and Waters [20] introduced a new concept for encryption, in which the sharing or access rules are expressed in the encryption algorithm itself. Essentially, the data owner provides a predicate or function $f$ describing how the data is to be accessed. Each user $u_i$ (where $i$ is an integer that indicates a specific user) possesses a set of credentials $\mathcal{A}_{u_i}$ out of a set of credentials $\mathcal{A}$. Then, the user with credentials $\mathcal{A}_{u_i}$ is able to decrypt a ciphertext with an expressed function $f$ if $f(\mathcal{A}_{u_i}) = 1$. In [20], the authors depicted a specific instance of this problem, hence called Attribute-Based Encryption (ABE). In ABE, a set of "attributes" represents a user's set of credentials, and a formula with these attributes as input represents the function. Essentially, the formula acts as an access policy.

This work was further expanded by [21], in which ABE was formulated in two complementary forms, Key-Policy ABE (KP-ABE) and Ciphertext-Policy ABE (CP-ABE). In the former, the ciphertexts are generated by including the desired complete set of attributes; the formulas over these attributes, or the access policies, are expressed in the users' secret keys. The latter prescribes that the attributes pertaining to a user are expressed in that user's secret key, whereas the formula or access policy is attached to the ciphertext at the moment of encryption [22]. Attribute-based Encryption has motivated numerous applications [23], [24], [25], [26], [27], [28],[29]. In this work, we propose to use CP-ABE for safe firearms operations.

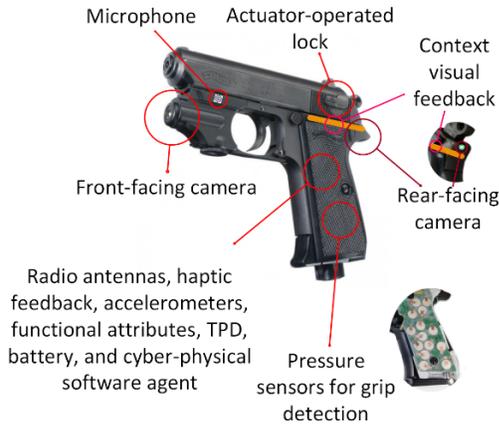

Fig. 1: Firearm model.

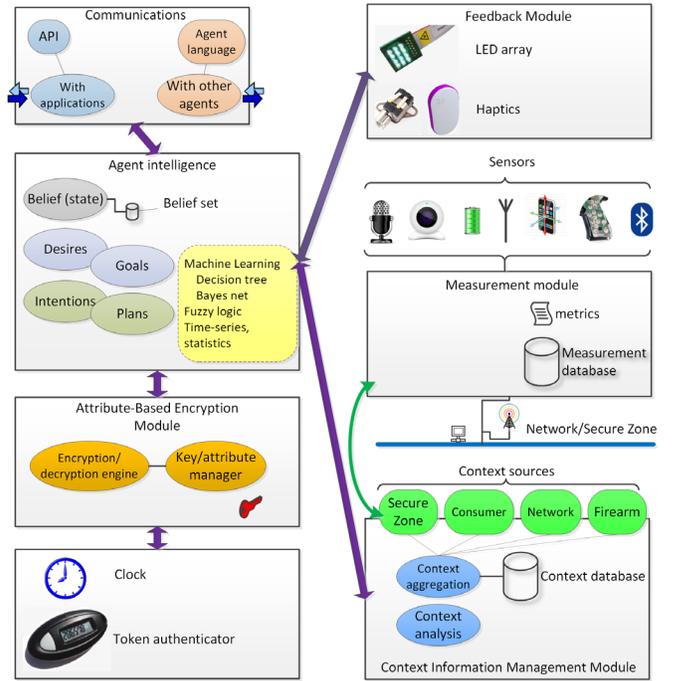

Fig. 2: Context-aware Agent Framework.

## VI. THE SECURE ZONE

The idea of areas in which guns are unsafe to be operated is certainly not new, and forms of it are in fact already in effect. Guns are generally not allowed beyond security checkpoints in airports, onboard airplanes, in most consulates and embassies, etc. These zones must be imposed by regulation and enforcement. We propose to address these safety areas within the firearm itself, with the aid of digital technologies.

One naïve approach to realizing safety areas would be to have the firearm communicate with a transmitter within the safety area and negotiate its operations. This operation would require two-way radio communications, which is still costly for devices that depend on battery, as it would be the case for the firearm in this example. Moreover, the two-way communications must employ secure and reliable protocols that might result in increased complexity and overhead.

The proper identification of the firearm or consumer poses another challenge. If a gun or consumer is to be uniquely identified to establish the authorization to operate the firearm, then a gun must transmit this identification to the area controller, which would then verify the authorization against a database of identifications. The area controller would then inform its authorization decision to the firearm. Although feasible, this protocol might incur scalability issues. Moreover, it would require two-way communications between the gun and the area controller, entailing a transmitter in the gun with sufficient power, and this would raise the energy (i.e., battery) requirements for the gun electronic hardware. We visualize a simpler, safer schema, which we depict in Fig. 3.

In this schema, a Secure Zone comprises one or more radio transmitters that will transmit a digital message containing the zone authorization for operation policy for firearms. The message consists of an *encrypted digital signature* together with the codification of a firearm operation policy for the zone, using CP-ABE. A known signature from a trusted authority and a private component, which is only known by the trusted authority, compose the encrypted digital signature. Each Secure Zone will have its own wireless transmission with its own encrypted message representing the firearm operation policy within that zone.

Individual firearms receive, at the time of purchase or during a registration process, a private key that encodes attributes of that firearm and those of the authorized consumer. For instance, a hunter would have attributes such as "OPEN_HUNTING_AREA, SHOOTING_RANGE". A security guard located at a court of justice would have attributes "COURT_JUSTICE, SHOOTING_RANGE". A civilian without special clearance could possess attribute "SHOOTING_RANGE" only, and a law-enforcement agent with high security clearance would retain a set of many attributes, or one attribute such as "LAW_ENFORCEMENT". (We propose and describe an infrastructure for key management and distribution tailored for our Secure Zones in Section VII.)

A firearm within a Secure Zone range receives the zone wireless broadcast transmission through its radio interface. The embedded software agent attempts to decode the encrypted message using the firearm's integrated private key, which contains the codified attributes of the firearm and its owner. If the agent successfully decrypts the message, it means that firearm is able to operate within that zone; otherwise, the operation is not safe, and the firearm's feedback components will actuate to inform the consumer of such. The haptic feedback component may vibrate, and the visual feedback component will illuminate to alert the consumer.

## VII. KEY INFRASTRUCTURE SCHEME

Our Secure Zone scheme uses CP-ABE, and encompasses message encryption, key generation, and key distribution. In this section, we compose an infrastructure to support our scheme, which deals with the specifics of our firearm conceptual design and Secure Zone scheme. This infrastructure builds upon systems proposed by [29], [30], [31]. In the following descriptions, we do not mention some components: further details can be found in the respective references.

### A. System Model

In our model, there are Secure Zone Authorities, a Central Authority, and consumers/firearms, depicted in Fig. 4. Each is described next.

*1) Central Authority (CA)*

The CA is managed by the government (the same branch that regulates firearms), and it is trusted. The CA's responsibility is to issue keys to firearms, manage the attribute set $\mathcal{A}$, and register Secure Zone Authorities, sending them cryptography parameters.

*2) Secure Zone Authority (SZA)*

The SZA is responsible for encrypting and wirelessly transmitting Secure Zone messages to the consumers/firearms. An SZA is managed by a zone administrator or owner, i.e., an SZA comprising a certain shopping mall is managed by that shopping mall's administration. SZA's must register with the CA to receive proper cryptography parameters.

*3) Firearms (F)*

A firearm receives its ABE secret key and other security parameters upon registration with the proper government branch. This secret key expresses the attributes established by the government, according to specific rules that consider the firearm and the consumer/owner.

### B. System Algorithms

A summary of the system algorithms follows.

**Setup**: randomized algorithm run by the CA and SZA. CA: takes the security parameter as input; outputs an ABE system public key and an ABE system master secret key; also generates a public/private key pair for CA usage. SZA: generates a public/private key pair for SZA usage.

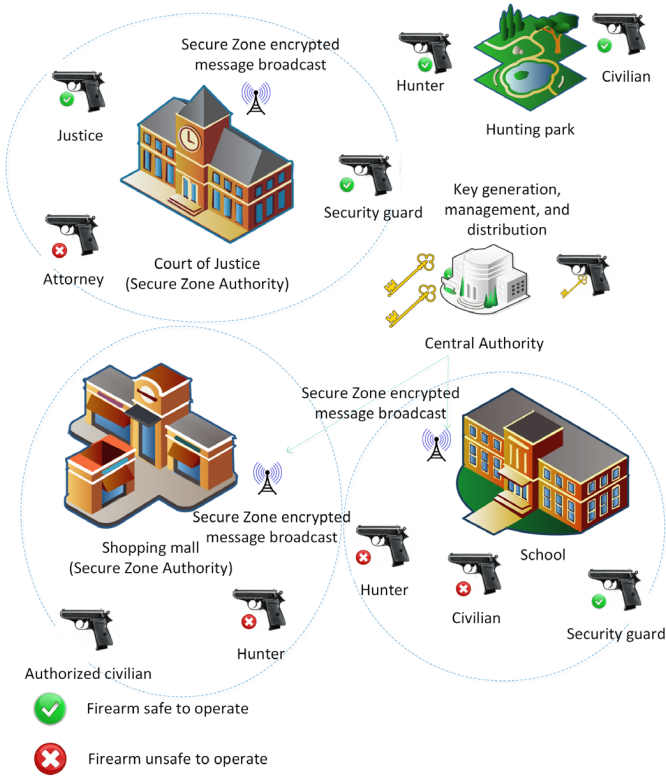

Fig. 3: Secure Zones.

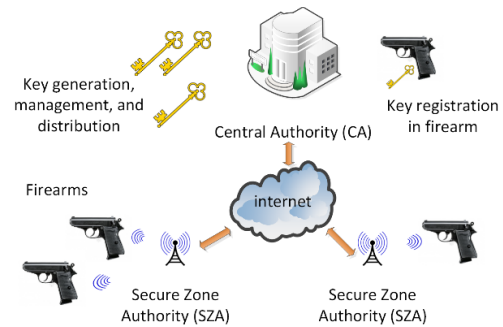

Fig. 4: System model.

**Attribute key generation:** randomized algorithm run by the CA. Input: CA's system master secret and public keys, a firearm's unique ID, a user's unique ID, a random number $x$, an expiration time $et$, and a set of attributes $\mathcal{A}_{u_i}$ possessed by the user/firearm $u_i$. Outputs user (ABE) secret key for the firearm.

**Token authenticator:** a function that generates a number $tk$ (the token) according to a seed and a message. The seed is a unique (random) secret, and the message is a certain number of periods elapsed since a given timestamp (such as the Unix epoch) [32], [33]. The idea is that if two parties have the same algorithm and the same seed (and synchronized clocks), legally obtained from a trusted source, both parties can generate the same number and thus one party can verify the authenticity of the other just by asking for the current generated number.

**Encryption:** randomized algorithm run by an SZA. Inputs: the Secure Zone policy for the SZA (i.e., the function $f$ as explained in Section V), a message $m$, a token $tk$, the SZA public key encrypted with the CA private key, a timestamp $ts$, and the ABE system public key. Outputs the encrypted ciphertext.

**Decryption:** deterministic algorithm run by the firearm software agent. Takes as input a ciphertext and a decryption key (the firearm/user ABE secret key), generated for an attribute set $\mathcal{A}_{u_i}$. Outputs the decrypted ciphertext if $f(\mathcal{A}_{u_i}) = 1$.

*C. Scheme Construction and Operation*

We describe how our infrastructure operates; some steps are not referenced and others, simplified.

*1) System Setup*

**Central Authority**: generates its ABE system public/master secret key. Generates its own public/private key. Securely stores each SZA's public key and securely sends this SZA public key back to the SZA, encrypted with the CA private key, generating $CA_{PK}(SZA_{PubK}^i)$, where $i$ is the particular SZA (the purpose of this will be explained later). Also sends along the system public key and the token authenticator algorithm and its system seed.

**Safe Zone Authority**: SZA registers with CA, receives a set of attributes $\mathcal{A}$. Generates its public/private keys. Securely sends its public key to CA, receives the system public key and $CA_{PK}(SZA_{PubK}^i)$ from the CA.

*2) Consumer Registration and Secret Key Generation*

Upon registration and consumer, or user, authentication with the government branch, a set of attributes $\mathcal{A}_u$ will be determined for the user $u$. The CA will generate the firearm/user ABE secret key expressing the user attributes (bonded by a random number $x$ to prevent colluding) together with the firearm ID, user ID, $x$ and expiration time $et$. This key is recorded into the firearm's TPD, along with the token authenticator algorithm, its system seed, and the CA's public key.

*3) Secure Zone Authority Encrypted Transmission*

$SZA^i$ decides on a Secure Zone policy $f$ over the attribute set $\mathcal{A}$. Generates token $tk$ and hashes it using standard hashing algorithm; obtains $hash(tk)$. Encrypts $hash(tk)$ with the SZA private key, generating $SZA_{PK}^i(hash(tk))$. Uses the token $tk$ as symmetric key to encrypt $SZA_{PK}^i(hash(tk))$ together with $CA_{PK}(SZA_{PubK}^i)$, and a timestamp $ts$. Uses the ABE system public key to encrypt previous result, $tk(SZA_{PK}^i(hash(tk)), CA_{PK}(SZA_{PubK}^i), ts)$, per the policy $f$; the final result is the message $m_{SZA^i}$. This message is broadcast via radio following a proper wireless communications standard.

*4) Decryption, Secure Operation Assessment*

A firearm $u_i$ within the range of $SZA^i$ Secure Zone receives, via its antenna, the message $m_{SZA^i}$. The software agent applies the decryption secret key. If the firearm has the appropriate attribute set $\mathcal{A}_{u_i}$ such that $f(\mathcal{A}_{u_i}) = 1$, then the decryption is successful, and the agent obtains $tk(SZA_{PK}^i(hash(tk)), CA_{PK}(SZA_{PubK}^i), ts)$. Agent runs its own token authenticator algorithm (with the same seed as the SZA, recorded at registration), obtains $tk_u$. Agent uses $tk_u$ as symmetric key to decrypt last outcome and get $ts$, $SZA_{PK}^i(hash(tk))$, and $CA_{PK}(SZA_{PubK}^i)$. If $tk_u \neq tk$, decrypt will fail; stop and alert user. Else, if $et < ts$ (secret key expired), then stop and alert user. Else, use the CA public key stored with the agent to decrypt $CA_{PK}(SZA_{PubK}^i)$. Use the resulting $SZA_{PubK}^i$ to decrypt $SZA_{PK}^i(hash(tk))$; if final product is different than $hash(tk_u)$ (apply the same standard hashing algorithm to $tk_u$, compare to $hash(tk)$), then at least one of the previous keys is invalid, or the message is invalid. Stop and alert user. Else, firearm is safe to operate within this Secure Zone. If, in the first step, $f(\mathcal{A}_{u_i}) \neq 1$, then decryption is unsuccessful, resulting in a message of invalid format. Stop and alert user.

By using a token authenticator, the software agent is able to detect problems with the transmission or a potential replay attack. If an attacker records a transmission and later replays it, there will be a token mismatch, as the token authenticator algorithm will generate different numbers based on the current time. Therefore, the hash of the token and the symmetric key (the token itself) will change periodically. Notice that the actual token is never transmitted; both parties (firearm and SZA) can generate similar tokens given the same secret seed, algorithm and current time.

Likewise, the simple timestamp and expiration time protocol allows the agent to verify the validity of the secret key. The agent can be programmed to enforce a key update as soon as an expiration event happens.

*5) Revocation and Expiration*

In this key infrastructure scheme, the key revocation happens when they expire. On this event, CA must generate a new consumer secret key, as described before.

*6) Key Update*

If a consumer wants to claim more attributes in addition to his/her current ones, follow the same procedure for consumer secret key generation, when CA will generate a new secret key.

## VIII. Conclusion and Future Work

We proposed Secure Zones, wireless-delimited areas within which encrypted messages using the Ciphertext-Policy Attribute-Based Encryption and expressing a zone's security policy are transmitted to our model application, firearms, within range. This cryptography maintains both privacy and security of communications, and its expressiveness means only firearms possessing the necessary attributes are able to decrypt those messages (through their agents). Thus, these firearms/agents can evaluate whether the firearm may operate within the zone. To support Secure Zones, we propose a Key Infrastructure for key generation, management, and distribution. The scheme uses a Central Authority for key generation, Secure Zone Authorities for encrypted message transmission, and timestamps and token authenticator algorithm for verifying authenticity and validity.

Furthermore, we presented our view of the firearm as a cyber-physical system. In our model, digital technologies enhance a firearm's safety through several components. In one component, we designed a conceptual model of a firearm fitted with an array of sensors and a context-aware software agent. The software agent uses the sensors and its agent programming to evaluate context and advise the consumer of potential unsafe operations. We introduced a Context-Aware Agent Framework, upon which those agents are built. Our initial experimentation focuses on implementing the key infrastructure and performing simulations, in which we evaluate and demonstrate whether mobile users within secure zones range can properly receive and decrypt the messages, if they have the appropriate attributes.

Our future work involves investigating the effectiveness of the proposed security model, and how to best address the issues of key revocation and update, as well as its performance in light of several attack vectors. Moreover, we will augment our model with delegating key generation to the Secure Zone Authorities, and prototype components of the system. We will scrutinize a suitable wireless communication infrastructure for the Secure Zones, such as [34] and [35]

REFERENCES

[1] I. Moir and A. G. Seabridge, Aircraft systems: mechanical, electrical, and avionics subsystems integration, 3 ed. Chichester, West Sussex, England; Hoboken, N.J.: John Wiley & Sons, Ltd, 2008.

[2] "The changing face of F1," ed: http://news.bbc.co.uk/sport2/hi/motorsport/formula_one/cars_guide/4272031.stm, 2005.

[3] K. Streams, "Google expects its self-driving cars to be ready in three to five years," ed: http://www.theverge.com/2013/2/11/3975988/google-expects-its-self-driving-cars-in-three-to-five-years, 2013.

[4] "Robotic surgery," ed: http://www.nlm.nih.gov/medlineplus/ency/article/007339.htm, 2011.

[5] Lindsey, "FDA takes fresh look at robotic surgery," ed: http://www.usatoday.com/story/news/nation/2013/04/09/robot-surgery-fda/2067629/, 2013.

[6] M. Gannon, "Exoskeleton allows paraplegics to walk," ed: http://www.cnn.com/2013/03/13/tech/innovation/original-ideas-exoskeleton, 2013.

[7] J. Cote, "Sandy Hook parents turn to tech for help," ed: http://www.sfgate.com/default/article/Sandy-Hook-parents-turn-to-tech-for-help-4356002.php, 2013.

[8] J. Johnson, "Can computer professionals and digital technology engineers help reduce gun violence?," Commun. ACM, vol. 56, pp. 35-37, March 2013.

[9] T. Associated Press, "Debate over futuristic 'smart gun' technology resumes as gun control issue takes center stage in national politics," ed: http://www.nydailynews.com/news/national/futuristic-smart-gun-technology-gun-control-solution-threat-2nd-amendment-article-1.1249466, 2013.

[10] "iGUN," ed: http://www.iguntech.com/, 2013.

[11] "TriggerSmart," ed: http://www.triggersmart.com/, 2013.

[12] "Dash7," ed: http://www.dash7.org/.

[13] "Armatix," ed: http://www.armatix.us/, 2013.

[14] D. Bobkoff, "Can 'Smart Gun' Technology Help Prevent Violence?," ed: NPR.org, 2013.

[15] Njit, "Spotlight: Smart Gun Technology Works," ed: http://www.njit.edu/news/spotlight/2005/jan/index.php, 2005.

[16] D. D'Orazio, "Galaxy S4 intros even more 'natural interactions' with eye tracking, gestures, and hover previews," ed: http://www.theverge.com/2013/3/14/4105798/galaxy-s4-software-features-air-view-smart-stay-announced, 2013.

[17] X. Liu, Z. Fang, and L. Shi, "Securing Vehicular Ad Hoc Networks," in Pervasive Computing and Applications, 2007. ICPCA 2007. 2nd International Conference on, 2007, pp. 424-429.

[18] C. Zhang, R. Lu, X. Lin, P.-H. Ho, and X. Shen, "An Efficient Identity-Based Batch Verification Scheme for Vehicular Sensor Networks," in INFOCOM 2008. The 27th Conference on Computer Communications. IEEE, 2008, pp. 246-250.

[19] M. Wooldridge, "Agent-based software engineering," Software Engineering. IEE Proceedings- [see also Software, IEE Proceedings], vol. 144, pp. 26 -37, February 1997.

[20] A. Sahai and B. Waters, "Fuzzy identity-based encryption," in Proceedings of the 24th annual international conference on Theory and Applications of Cryptographic Techniques, Berlin, Heidelberg, 2005, pp. 457-473.

[21] V. Goyal, O. Pandey, A. Sahai, and B. Waters, "Attribute-based encryption for fine-grained access control of encrypted data," in Proceedings of the 13th ACM conference on Computer and communications security, New York, NY, USA, 2006, pp. 89-98.

[22] B. Waters, "Ciphertext-policy attribute-based encryption: an expressive, efficient, and provably secure realization," in Proceedings of the 14th international conference on Practice and theory in public key cryptography conference on Public key cryptography, Berlin, Heidelberg, 2011, pp. 53-70.

[23] N. Chen, M. Gerla, D. Huang, and X. Hong, "Secure, selective group broadcast in vehicular networks using dynamic attribute based encryption," in Ad Hoc Networking Workshop (Med-Hoc-Net), 2010 The 9th IFIP Annual Mediterranean, 2010, pp. 1-8.

[24] X. Hong, D. Huang, M. Gerla, and Z. Cao, "SAT: situation-aware trust architecture for vehicular networks," in Proceedings of the 3rd international workshop on Mobility in the evolving internet architecture, New York, NY, USA, 2008, pp. 31-36.

[25] D. Huang, X. Hong, and M. Gerla, "Situation-aware trust architecture for vehicular networks," Communications Magazine, IEEE, vol. 48, pp. 128-135, 2010.

[26] L. Ibraimi, M. Asim, and M. Petkovic, "Secure management of personal health records by applying attribute-based encryption," in Wearable Micro and Nano Technologies for Personalized Health (pHealth), 2009 6th International Workshop on, 2009, pp. 71-74.

[27] X. Liang, Z. Cao, H. Lin, and J. Shao, "Attribute based proxy re-encryption with delegating capabilities," in Proceedings of the 4th International Symposium on Information, Computer, and Communications Security, New York, NY, USA, 2009, pp. 276-286.

[28] S. Luo, J. Hu, and Z. Chen, "Implementing Attribute-Based Encryption in Web Services," in Web Services (ICWS), 2010 IEEE International Conference on, 2010, pp. 658-659.

[29] J. Hur and K. Kang, "Secure Data Retrieval for Decentralized Disruption-Tolerant Military Networks," IEEE/ACM Transactions on Networking (to appear), vol. PP, 2012.

[30] L. Yeh and J. Huang, "PBS: A Portable Billing Scheme with Fine-Grained Access Control for Service-Oriented Vehicular Networks," Mobile Computing (to appear), IEEE Transactions on, vol. PP, 2013.

[31] M. Chase, "Multi-authority attribute based encryption," in Proceedings of the 4th conference on Theory of cryptography, Berlin, Heidelberg, 2007, pp. 515-534.

[32] D. M'Raihi, M. Bellare, F. Hoornaert, D. Naccache, and O. Ranen, "HOTP: An HMAC-Based One-Time Password Algorithm," ed: IETF RFC 4226, 2005.

[33] D. M'Raihi, S. Machani, M. Pei, and J. Rydell, "TOTP: Time-Based One-Time Password Algorithm," ed: IETF RFC 6238, 2011.

[34] M. Yun, D. Kim, H. seok Lee, and J. Lee, "Silent broadcast: Experience of connectionless messaging Using Wi-Fi P2P," in Information Science and Digital Content Technology (ICIDT), 2012 8th International Conference on, 2012, pp. 239-242.

[35] S. S. Chawathe, "Low-latency indoor localization using bluetooth beacons," in Intelligent Transportation Systems, 2009. ITSC '09. 12th International IEEE Conference on, 2009, pp. 1-7.